# Dynamical analogy between economical crisis and earthquake dynamics within the nonextensive statistical mechanics framework

Stelios M. Potirakis[a], Pavlos I. Zitis [b], and Konstantinos Eftaxias[b]

*a. Department of Electronics Engineering, Technological Education Institute (TEI) of Piraeus, 250 Thivon & P. Ralli, GR-12244, Aigaleo, Athens, Greece, spoti@teipir.gr .*
*b. Department of Physics, Section of Solid State Physics, University of Athens, Panepistimiopolis, GR-15784, Zografos, Athens, Greece, ceftax@phys.uoa.gr .*

**Abstract**
The field of study of complex systems considers that the dynamics of complex systems are founded on universal principles that may be used to describe a great variety of scientific and technological approaches of different types of natural, artificial, and social systems. Several authors have suggested that earthquake dynamics and the dynamics of economic (financial) systems can be analyzed within similar mathematical frameworks. We apply concepts of the nonextensive statistical physics, on time-series data of observable manifestations of the underlying complex processes ending up to these different extreme events, in order to support the suggestion that a dynamical analogy exists between a financial crisis (in the form of share or index price collapse) and a single earthquake. We also investigate the existence of such an analogy by means of scale-free statistics (the Gutenberg–Richter distribution of event sizes). We show that the populations of: (i) fracto-electromagnetic events rooted in the activation of a single fault, emerging prior to a significant earthquake, (ii) the trade volume events of different shares / economic indices, prior to a collapse, and (iii) the price fluctuation (considered as the difference of maximum minus minimum price within a day) events of different shares / economic indices, prior to a collapse, follow both the traditional Gutenberg–Richter law as well as a nonextensive model for earthquake dynamics, with similar parameter values. The obtained results imply the existence of a dynamic analogy between earthquakes and economic crises, which moreover follow the dynamics of seizures, magnetic storms and solar flares.

## 1. Introduction

The relatively new field of complex systems increasingly gains the interest of scientists working on disciplines ranging from physics and engineering to economics, biosciences and social sciences, e.g., [1-9]. It has been suggested that a unique characteristic of complex systems, which is really intriguing, is that their dynamics are probably governed by a set of universal principles that may be used to describe a great variety of different natural, artificial, and social systems [2].

Especially for systems exhibiting long-range correlations, memory, or fractal properties, as the complex systems, nonextensive statistics becomes an appropriate mathematical tool [10-12] in order to examine the suggestion of the existence of a common approach to the interpretation of different complex extreme events in terms of the same driving dynamical system mechanism. Herein, we investigate the aforementioned suggestion in terms of nonextensive statistical mechanics focusing on two different complex extreme events, namely, seismic shocks and financial crashes.

A central property of the earthquake (EQ) preparation process is the occurrence of coherent large-scale collective behavior with a very rich structure, resulting from repeated nonlinear interactions among the constituents, namely, opening cracks, of the system [13,14]. An opening crack behaves as an efficient electromagnetic (EM) emitter. Consequently, nonextensive statistics is also a physically meaningful tool for investigating the launch of a fracture induced EM precursors, since these are observable manifestations of the underlying EQ preparation process, e.g., [15-17]. Features analogous to the ones appearing in time-series originating from the EQ preparation process, like the seismogenic EM emissions, are also present in time-series that are observable manifestations of other complex systems approaching crisis. Dynamical analogies between seismogenic EM emissions and signals from biological systems driven to extreme operation [18-22], as well as signals used for the observation of solar flares and magnetic storms [23-27] have already been pointed out, most of them within the frame of nonextensive statistical mechanics [21,24-28].

Perhaps one of the most vivid and richest examples of the dynamics of a complex system at work is that of economic systems (financial markets) [29-38]. Their richness in interactions renders them characteristic examples of complex dynamics. Nonextensive characteristics have already been investigated on economic time-series, yielding that the economic systems are in most cases well described by nonextensive statistics, e.g., [38-46].

The identification of common characteristics /laws that govern the dynamics of systems as different as the preparation process of an EQ and the economic process which leads to the collapse of the price of a share or an economic index seems to be very interesting; the exchange of information between these different fields is to the benefit of both of them. If, for example, dynamical characteristics associated with an increased possibility for a significant EQ to occur can be identified in an economic time-series, probably a better estimation of financial risk may be obtained.

Complex systems seem to occur close, in some sense, to the frontier between order and disorder where most of their basic quantities exhibit nonexponential behaviors, very frequently power laws. It happens that the distributions and other relevant quantities that emerge naturally within the frame of nonextensive statistical mechanics are precisely of this type [1]. As it was mentioned, herein our analysis is mainly focused on the nonextensive Tsallis statistical mechanics, namely, a recently introduced nonextensive model for EQ dynamics [47]. One universal footprint seen in many complex systems is self-affinity and the fractional power law relationship is a typical expression of a self-affine structure - the fractal. Note that the nonextensive model also leads to a fractional power law relationship [48,49].

Extending a recently proposed modeling of market index returns in terms of the well known Gutenberg-Richter (G-R) law for the EQ dynamics [50], we perform our analysis mainly based on a recently introduced nonextensive model for EQ dynamics which leads to a G-R type law for the relationship between frequency and magnitude of EQs [47,51]. We show that the populations of: (i) fracto-EM events rooted in the activation of a single fault, emerging prior to a significant EQ, (ii) the trade volume events of different shares / economic indices, prior to a collapse, and (iii) the price fluctuation (considered as the difference of maximum minus minimum price within a day) events of different shares / economic indices, prior to a collapse, follow the above mentioned nonextensive statistical law as well as the traditional G-R law, namely exhibit similar scaling properties, with similar non-extensive $q-$ parameter, i.e, imply the existence of universality.

The remaining of this contribution is organized as follows: In Sec. 2, we discuss in brief the notion of seismogenic EM emissions associated with the occurrence of an earthquake and how the EM data are taken. A short introduction to nonextensive statistics is provided in Sec. 3. The necessary background information on the G-R frequency-size law, as well as on a nonextensive frequency-size model of EQ dynamics is introduced in Sec. 4. In Sec. 5, the analysis of different economic time-series is presented and the obtained results are discussed in comparison to already published results for EM seismogenic time-series, which were obtained using the same analysis methods. Finally, the conclusions are discussed in Sec. 6.

## 2. The notion of seismogenic EM emissions

Earthquake is mainly a large-scale dynamic failure process. A geological fault is comprised of two rock surfaces in contact. Accumulated evidence shows that a fault surface is a fractal object. Consequently, a fault may be viewed as pair of overlapping fractals. An earthquake occurs when one fractal profile is drifting on another [52, and references therein].

More precisely, the way in which a frictional interface fails is crucial to our fundamental understanding of failure processes. The essential microscopic ingredient in friction is asperity. A magnified view of fault surfaces reveals a rough looking surface with high asperities and low valleys. The two macroscopic surfaces in contact are indeed detached almost everywhere except for asperities. But, how does the family of asperities break? The following model is widely accepted [53, and references therein]. A large EQ initiates at a small nucleation area and grows as propagating rupture fronts. The propagating fronts activate a multitude of fault patches (asperities) that undergo intense deformation (see Fig. 1 in [53]). Before the front arrives, the stress on each patch is generally lower than its static strength. If the arriving front raises the stress to the static strength level, the patch strength may drop and it slips releasing elastic energy stored in the rocks, then eventually is decelerates and stops. In this way, the frictional fault surfaces suddenly slip, lock and then slip again in a repetitive manner forming the *"stick-slip"* state. The repetition of such local damage-slip events intensifies fault wear and dynamic weakening. The discovery of slow-slip phenomena has revolutionized our understanding of how faults accommodate relative plate motions [54].

Crack generation and propagation is inherent to brittle failure, while opening crack dynamics comprises the basis of all theories attempting to describe the processes leading to failure. An opening crack produces EM emission. Fracture-induced EM emissions are sensitive to the structural changes. Thus, their study constitutes a basic method for the monitoring of the evolution of damage process both on the laboratory and the geophysical scale. The origin of EM emission from fracture can be explained in the frame of the "capacitor model" [52, and references therein]. More precisely, in many materials, emission of photons, electrons, ions and neutral particles are observed during the formation of new surface. The rupture of inter-atomic (ionic) bonds leads to intense charge separation, which is the origin of the electric charge between the crack faces. On the faces of a newly created micro-crack the electric charges constitute an electric dipole or a more complicated system. The motion of a crack has been shown to be governed by a dynamical instability causing oscillations in its velocity and structure of the fractured surfaces. It is worth mentioning that laboratory experiments show that more intense fracto-emissions are observed during the unstable crack growth. Due to the crack strong wall vibration, in the stage of the micro-branching instability, the opening crack behaves as an efficient EM emitter [52, and references therein]. Based on the above mentioned generation mechanism of fracture induced EM emission it is reasonable to accept that during the damage of an asperity an EM avalanche (a fracto-EM emission event or "*electromagnetic earthquake" (EM-EQ)) is emitted.* Accumulated experimental evidence suggest that the kilohertz (kHz) EM emission observed in the field is rooted in the fracture of asperities distributed along a fault sustaining the system, e.g., [55-57, and references therein]. *In the frame of this proposal, the sequence of "EM-EQs" included in an emerged avalanche-like strong kHz EM emission associated with the activation of a single fault mirrors the sequential damage of asperities, i.e., the sequence of the "stick-slip" events that characterize the stage of quasi-static stick-slip-like slow sliding.* The number of "electromagnetic EQs" included in the kHz EM emission may correspond to the fracture of critical number of strong patches that results in a dramatic fault wear and dynamic weakening, and thus to the start of the preparation of the final stage that results to the global fast slip, i.e. the EQ occurrence. Due to its crucial character, the above mentioned proposal has been supported by strong documentation [52, and references therein].

Now we focus on the notion of the "electromagnetic earthquake". It is defined as follows, e.g., [55-57]: As it was said, a single EM burst occurs when there is an intersection between the two fractal profiles of the fault. If $A(t_i)$ refers to the amplitude of the pre-EQ EM time-series, at time sample $t_i$, we regard as amplitude of a candidate "fracto-EM emission" the difference $A_{fem}(t_i) = A(t_i) - A_{noise}$, where $A_{noise}$ is the maximum value of the EM recording during a "quiet" period, namely far from the time of the EQ occurrence. We consider that a sequence of $k$ successively emerged "fracto-electromagnetic emissions" $A_{fem}(t_i)$, $i = 1,\ldots,k$ represents the EM energy released, $\varepsilon$, during the damage of a fragment. We shall refer to this as an "electromagnetic earthquake" (EM-EQ). Since the sum of the squared amplitude of the fracto-EM emissions is proportional to their energy, the magnitude $M$ of the candidate EM-EQ is given by the relation $M \propto \log \varepsilon = \log \left( \sum \left[ A_{fem}(t_i) \right]^2 \right)$.

*Data collection*: Since 1994, a station has been installed and operated at a mountainous site of Zante island (37.76° N–20.76° E) in the Ionian Sea (western Greece). The main aim of this station is the detection of fracture induced EM seismic precursors. Six loop antennas detect the three components (EW, NS, and vertical) of the variations of the magnetic field at 3 kHz and 10 kHz respectively. All the EM time series are sampled at 1 Hz. Clear kHz EM precursors have been detected over periods ranging from approximately a few days to a few hours prior to catastrophic EQs that occurred in Greece or Italy (e.g., [55-71]). Recent results indicate that the recorded EM precursors contain information characteristic of an ensuing seismic event (e.g., [55-71]).

3. Nonextensive statistics

In nature, long-range spatial interactions or long-range memory effects may give rise to very interesting behaviors. Among them, one of the most intriguing arises in systems that are nonextensive (nonadditive). These systems share a very subtle property: they violate the main hypothesis of Boltzmann-Gibbs (B-G) statistics, i.e. ergodicity. Inspired by multifractals concepts, Tsallis [1,10,11,72] has proposed a generalization of the B-G statistical mechanics, which covers systems that violate ergodicity, systems the microscopic configurations of which cannot be considered as (nearly) independent. This generalization is based on nonadditive entropies, $S_q$, characterized by an index $q$ which leads to a nonextensive statistics [73],

$$S_q = k \frac{1}{q-1} \left( 1 - \sum_{i=1}^{w} p_i^q \right), \qquad (1)$$

where $p_i$ are the probabilities associated with the microscopic configurations, $w$ is their total number, $q$ is a real number, and $k$ is Boltzmann's constant. The value of $q$ is a measure of the nonextensivity of the system. Notice, $q = 1$ corresponds to the standard, extensive, B-G statistics.

The entropic index $q$ characterizes the degree of nonextensivity reflected in the following pseudo-additivity rule:

$$\frac{S_q(A+B)}{k} = \frac{S_q(A)}{k} + \frac{S_q(B)}{k} + (q-1)\frac{S_q(A)}{k}\frac{S_q(B)}{k}. \tag{2}$$

For subsystems that have special probability correlations, additivity

$$S_{B-G}(A+B) = S_{B-G}(A) + S_{B-G}(B) \tag{3}$$

is not valid for $S_{B-G}$, but may Eq. (2) occur for $S_q$ with a particular value of the index $q$. Such systems are referred to as nonextensive [1,11].

The cases $q > 1$ and $q < 1$, correspond to sub-extensivity, or super-extensivity, respectively. We may think of $q$ as a bias-parameter: $q < 1$ privileges rare events, while $q > 1$ privileges prominent events [74].

We emphasize that the parameter $q$ itself is not a measure of the complexity of the system but measures the degree of nonextensivity of the system. It is the time variations of the Tsallis entropy, $S_q$, for a given $q$ that quantify the dynamic changes of the complexity of the system. Lower $S_q$ values characterize signals with lower complexity.

**4. Frequency-size laws**

Earthquake dynamics have been found to follow the frequency-magnitude scaling relation, known as Gutenberg - Richter (G-R) law [75]

$$\log N(>M) = a - bM, \tag{4}$$

where $N(>M)$ is the number of earthquakes with magnitude greater than $M$ occurring in a specified area and time and the coefficient $b$, called "the $b-$value", is the negative slope of $\log N(>M)$ vs. $M$ diagram; $a$ is a real constant.

A model for EQ dynamics based on a non-extensive Tsallis formalism, starting from fundamental principles, has been recently introduced by Sotolongo-Costa and Posadas [47] and revised by Silva et al. [51]. This approach leads to a non-extensive G-R type law for the magnitude distribution of EQs:

$$\log[N(>M)] = \log N + \left(\frac{2-q}{1-q}\right)\log\left[1 - \left(\frac{1-q}{2-q}\right)\left(\frac{10^{2M}}{a^{2/3}}\right)\right], \tag{5}$$

where $N$ is the total number of EQs, $N(>M)$ the number of EQs with magnitude larger than $M$, $M \propto \log \varepsilon$. $a$ is the constant of proportionality between the EQ energy, $\varepsilon$, and the size of fragment, $r$, $(\varepsilon \propto r^3)$. It is reminded that the entropic index $q$ characterizes the degree of nonextensivity. Importantly, the associated with Eq. (5) $q-$values for different regions (faults) in the world are restricted in the region 1.6 - 1.7 [51].

The $q-$parameter included in the nonextensive formula of Eq. (5), *above some magnitude threshold*, is associated with the $b-$value by the relation [48]:

$$b_{est} = 2 \cdot \frac{2-q}{q-1} \qquad (6)$$

## 5. Time-series analysis

In the following, the notion of an economic event on the field of economic time-series is introduced in a way analogous to the notion of fracto-EM event, or "EM-EQ" (see Sec. 2) for the fracto-EM time-series. Then, based on this definition for the economic event, specific shares and economic indices which presented crisis (sudden collapse of their price) are analyzed both using the G-R law (Eq. (4)) and the nonextensive model of Eq. (5). The results are discussed as to the ability of the two frequency-size equations to model the studied economic event series, as well as to their similarity to the results obtained for fracto-EM event series and foreshock seismicity either at the geophysical or the laboratory fracture experiments scale.

*5.1. Economic events in analogy to fracto-EM events*

Both G-R law and its non-extensive variant of Eq. (5) were initially proposed as seismicity models. Therefore, the notion of EQ magnitude is of central importance in these models. We remind that, although a number of different definitions of EQ magnitude do exist, all of them present a common characteristic, they are quantities proportional to some expression of EQ energy, i.e., $M \propto \log E$. In order for the frequency-size laws to be applied, a corresponding "magnitude" has to be defined. Based on the generation mechanism of fracto-EM emissions [52, and references therein] (see also Sec. 2), the notion of fracto-EM emission event, or "electromagnetic earthquake" (EM-EQ) (see Sec. 2), has been successfully proposed within the frame of the self-affine nature of fracture and faulting [15-17,55-57].

Although not having a clear physical meaning, the notion of economic event is proposed in direct analogy to the fracto-EM event. Of course, it is not possible to define energy, in the strict sense, on an economic time-series, since its values correspond to the values of a quantity without specific physical meaning / units. However, one can define signal's power as used in the signal processing sense, meaning generally a quantity, a function, being proportional to the squared signal. Within this frame, the magnitude of an economic event is proposed to be calculated as $M \propto \log \varepsilon = \log\left(\sum [A_{ec}(t_i)]^2\right)$, where $A_{ec}(t_i) = A(t_i) - A_{"noise"}$, with $A(t_i)$ representing the "amplitude" of the economic time-series (in our investigation the trade volume or the price fluctuation), and $A_{"noise"}$ representing the corresponding maximum value during a "quiet" period, namely far from the time of the crisis occurrence.

*5.2. Analysis of economic time-series prior to crisis*

We proceed now to the analysis of financial market observables, specifically traded volumes, from now on denoted by the capital letter $V$, and daily price fluctuations of different shares and economic indices, from now on denoted by the capital letter $F$. Note that as "daily price fluctuation" the difference of maximum minus minimum price within a day is considered, namely, $F(day_i) = \max\left(price_{day_i}\right) - \min\left(price_{day_i}\right)$, where $day_i$ is the $i-$th day, and $price_{day_i}$ represents the set of different prices of a specific share or economic index recorded within the $i-$th day. Accordingly, we shall refer to economic events resulting, the way described in SubSec. 5.1, from the traded volumes time-series as the "$V-$events", while the ones resulting from the daily price fluctuation time-series as the "$F-$events". The economic time-series data have been retrieved from the historical prices record of http://finance.yahoo.com/. Specifically, the following data have been studied; for each of them a wider time period was first examined to locate a crisis point (price collapse) and then a narrower excerpt of the time-series prior to the crisis was analyzed:

• the S&P/TSX Composite index (GSPTSE): from 3/1/2000 up to 3/8/2012; on 17/6/2008 the daily closing price reaches its highest price before it collapses, therefore the time period (3/1/2000 - 17/6/2008) was analyzed.
• the Hang Seng Index (HSI): from 5/5/2003 up to 5/5/2009; on 6/11/2007 the daily closing price reaches its highest price before it collapses, therefore the time period (5/5/2003 - 6/11/2007) was analyzed.

- the index NASDAQ Composite (IXIC): from 5/2/1971 up to 9/11/2012; on 24/2/2000 the daily closing price reaches its highest price before it collapses, therefore the time period (5/2/1971 - 24/2/2000) was analyzed.
- the index S&P 500 (GSPC): from 3/1/1950 up to 9/11/2012; on 17/7/2000 the daily closing price reaches its highest price before it collapses, therefore the time period (3/1/1950 - 17/7/2000) was analyzed.
- the index SPDR Dow Jones Industrial Average (DIA): since 20/1/1998 up to 3/8/2012; on 15/10/2007 the daily closing price reaches its highest price before it collapses, therefore the time period (20/1/1998 - 15/10/2007) was analyzed.
- the share United States Steel Corp. (X): since 12/4/1991 up to 3/8/2012; on 11/6/2008 the daily closing price reaches its highest price before it collapses, therefore the time period (12/4/1991 - 11/6/2008) was analyzed.
- the share Potash Corp. of Saskatchewan, Inc. (POT): since 12/1/1990 up to 9/11/2012; on 23/6/2008 the daily closing price reaches its highest price before it collapses, therefore the time period (12/1/1990 - 23/6/2008) was analyzed.
- the share Carrizo Oil & Gas Inc. (CRZO): since 6/8/1997 up to 9/11/2012; on 21/5/2008 the daily closing price reaches its highest price before it collapses, therefore the time period (6/8/1997 - 21/5/2008) was analyzed.

As an example, Fig.1 shows daily closing price, $C$, as well as the trade volume, $V$, and the daily price fluctuation, $F$, for the Hang Seng Index (HSI), while Fig. 2 shows the corresponding time-series for the share Potash Corp. of Saskatchewan, Inc. (POT). Note that for the second case, the daily closing price adjusted for dividends and splits, $A.C.$, is shown as an indication for the time evolution of share price.

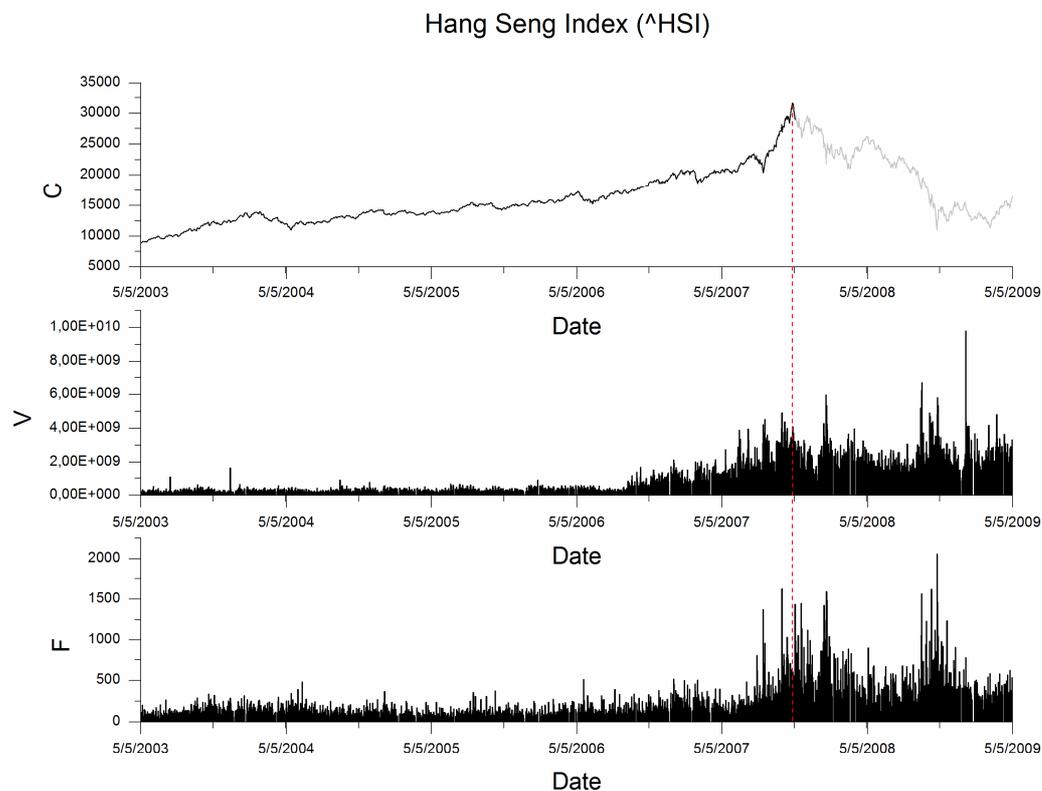

**Fig. 1.** The Hang Seng Index (HSI). Top graph depicts the daily closing price, $C$, as an indication for the time evolution of index's price. The intense black colored part of the time-series indicates the pre-crisis analyzed time period. The middle graph depicts the trade volume, $V$, and the bottom graph the daily price fluctuation, $F$, i.e. the difference between the highest and the lowest of the price for each day. The red line marks the date 6/11/2007 that the daily closing price reached its highest price before it collapses.

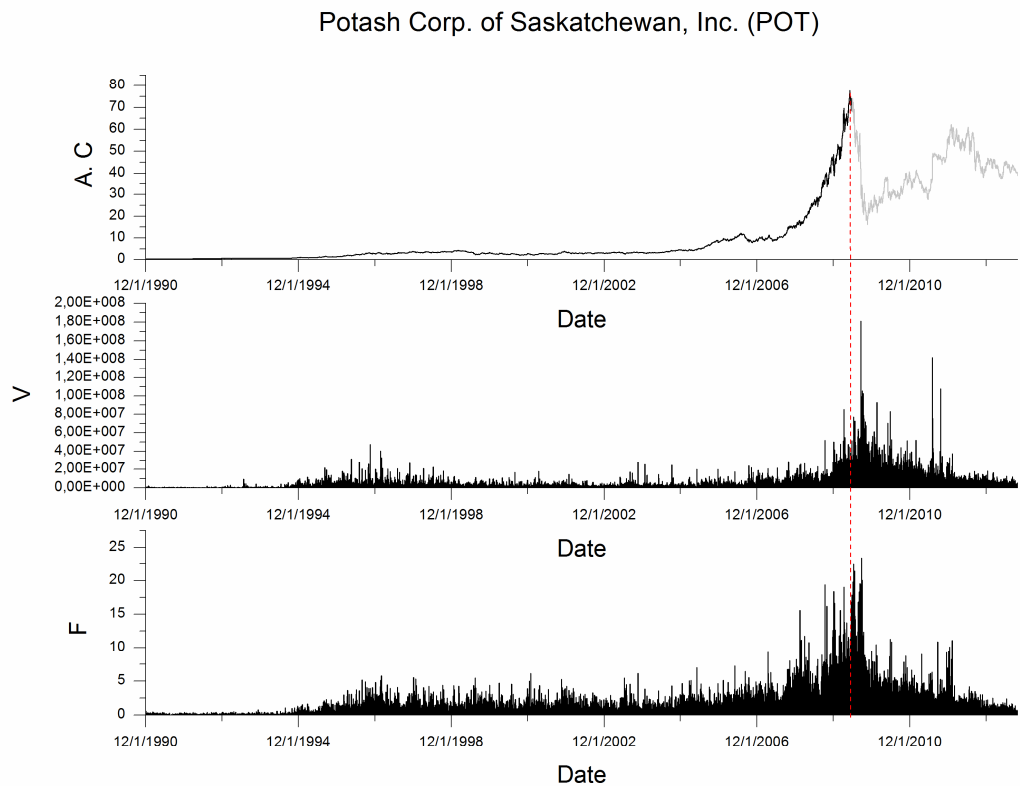

**Fig. 2.** The share Potash Corp. of Saskatchewan, Inc. (POT). Top graph depicts the daily closing price adjusted for dividends and splits, $A.C.$, as an indication for the time evolution of share price. The intense black colored part of the time-series indicates the pre-crisis analyzed time period. The middle graph depicts the volume of trade, $V$, and the bottom graph the daily price fluctuation, $F$, i.e. the difference between the highest and the lowest of the price for each day. The red line marks the date 23/6/2008 that the daily closing price reached its highest price before it collapses.

It is clarified at this point that the "quiet" period used for the determination of the parameter $A_{"noise"}$ is always selected far from the time of the crisis occurrence and specifically before the increasing parts of the analyzed time-series, in order not to lead to biased estimates of the magnitude $M$. For example, in the case of the Hand Seng Index (Fig. 1) the quiet period cannot include the increase which starts approximately one year before the crisis for the variable $V$ and about half-year before the crisis for the variable $F$. Actually, a time-period of ~ half-year (12/12/2002-24/4/2003) was selected as the quiet period in this case. Dependent on the case, the quiet period may be taken even within the analyzed period provided that, as already stressed, it is far from the time of the crisis occurrence, while the quiet periods can be different for the two time-series analyzed ($V$ and $F$). As soon as the quiet period is determined, the threshold values $A_{"noise"}$ for $V$ and $F$ are selected to correspond to a probability of ~ 90% that all the quiet period values are lower than the selected thresholds. This approach is considered to give a satisfactory estimation of the "maximum" value of the "noise" without the risk of an overestimation of the threshold value due to (very rare) outlier values. The threshold values $A_{"noise"}$ for each one the analyzed cases are given in the legends of the figures portraying the analysis results (Figs 3-10).

Both frequency-size laws were fitted in the time domain, on a single excerpt of each time-series. The results are shown in Figs 3-10, where $G(>M) = N(>M)/N_{tot.}$, with $N_{tot.}$ the total number of events. Also note that a magnitude normalization has been performed in order for the economic event for all the examined shares/ indices to start from an event magnitude value equal to zero, for a uniform / comparable representation; the normalization does not affect the estimation of the $q-$ values or the $b-$values.

From Figs 3-10 it can be concluded that, in general, the economic events are well represented either by the G-R law or the nonextensive model for EQ dynamics of Eq. (5). The economic data are well fitted, *above some* *magnitude threshold,* by the G-R law, while Eq. (5) is well fitted *for the whole range of events*. The obtained $q-$ values (resulting directly from fitting of Eq. (5)), the $b-$ values (resulting directly from fitting of G-R law, Eq. (4)), as well as the $b_{est}-$ values (resulting from the corresponding $q-$ values and Eq. (6)) are summarized in Table 1. For most of the cases (11 out of 16) there is a very close agreement, within the estimation tolerances, between the estimated, $b_{est}-$, and the directly calculated, $b-$, values for the eight different examples analyzed and the two different event definitions for each one of them, while for the ones that present differentiation this is clearly marginal. These results verify that: (i) the nonextensive model of Eq. (5) can successfully describe the analyzed economic data and (ii) that both event definitions (for $V$ and $F$) are appropriate for the investigation of the economic data dynamics. Note that the herein resulted $b-, b_{est}-$ values are also in agreement with the ones recently obtained for a different definition of the economic event magnitude and the analysis of a large number of US stock market crashes, based on daily data of the Dow Jones Industrial Index, the great crash of 1929 and the crash of 1987 included [50].

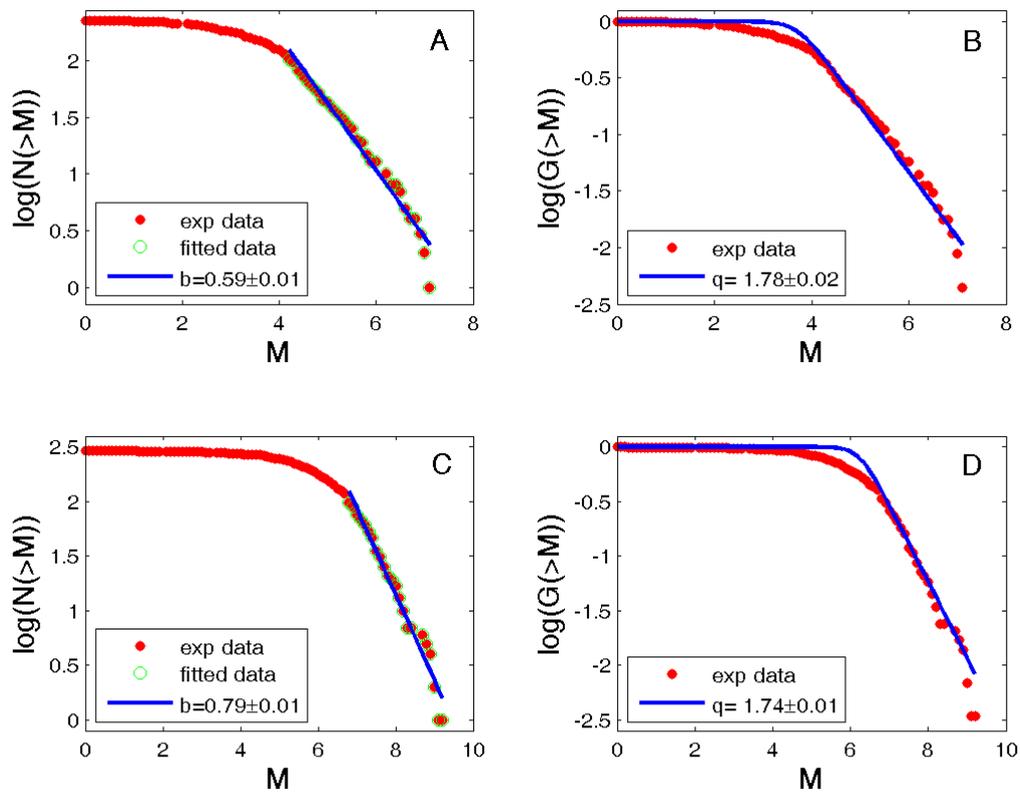

**Fig. 3.** The S&P/TSX Composite index (GSPTSE). A. Fitting of G-R law on the trade volume events; B. Fitting of the nonextensive formula of Eq. (5) on the trade volume events; C. Fitting of G-R law on the daily price fluctuation events; D. Fitting of the nonextensive formula of Eq. (5) on the daily price fluctuation events. The employed threshold values $A_{"noise"}$ for $V$ and $F$ were 1.10E+08 and 100, respectively.

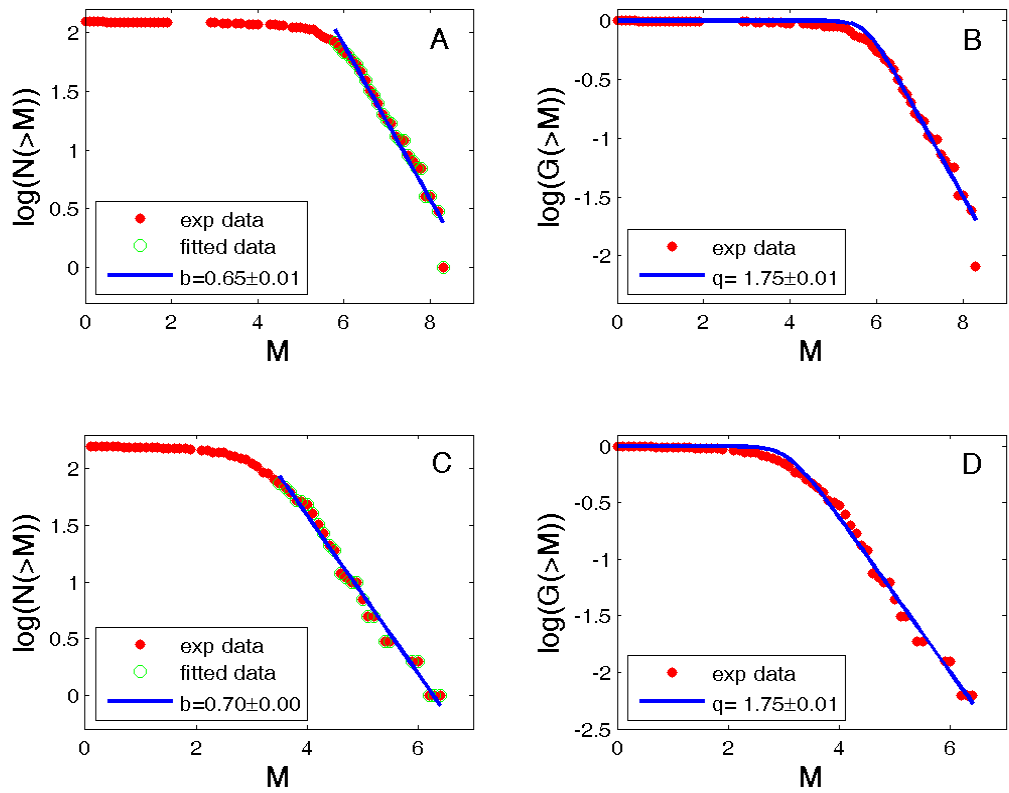

**Fig. 4.** The Hang Seng Index (HSI). A. Fitting of G-R law on the traded volume, $V-$, events; B. Fitting of the nonextensive formula of Eq. (5) on the traded volume, $V-$, events; C. Fitting of G-R law on the daily price fluctuation, $F-$, events; D. Fitting of the nonextensive formula of Eq. (5) on the daily price fluctuation, $F-$, events. The employed threshold values $A_{"noise"}$ for $V$ and $F$ were 2.70E+08 and 200, respectively.

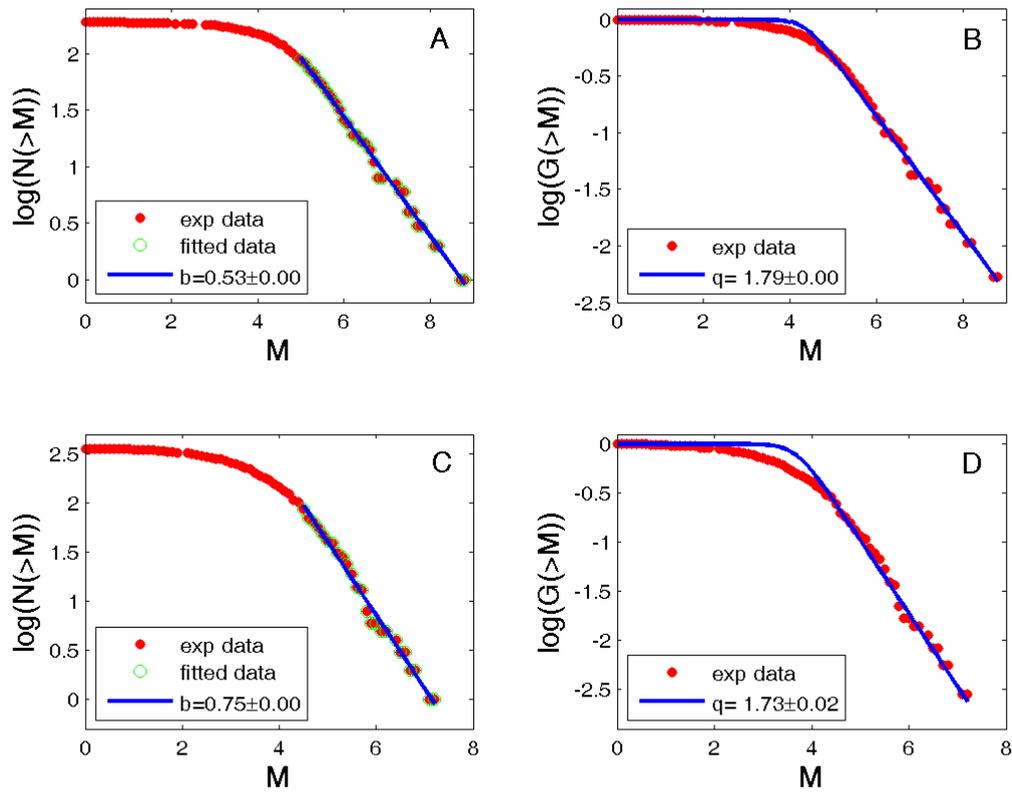

**Fig. 5.** The index NASDAQ Composite (IXIC). A. Fitting of G-R law on the traded volume, $V-$, events; B. Fitting of the nonextensive formula of Eq. (5) on the traded volume, $V-$, events; C. Fitting of G-R law on the daily price fluctuation, $F-$, events; D. Fitting of the nonextensive formula of Eq. (5) on the daily price fluctuation, $F-$, events. The employed threshold values $A_{"noise"}$ for $V$ and $F$ were 1.20E+08 and 4.0, respectively.

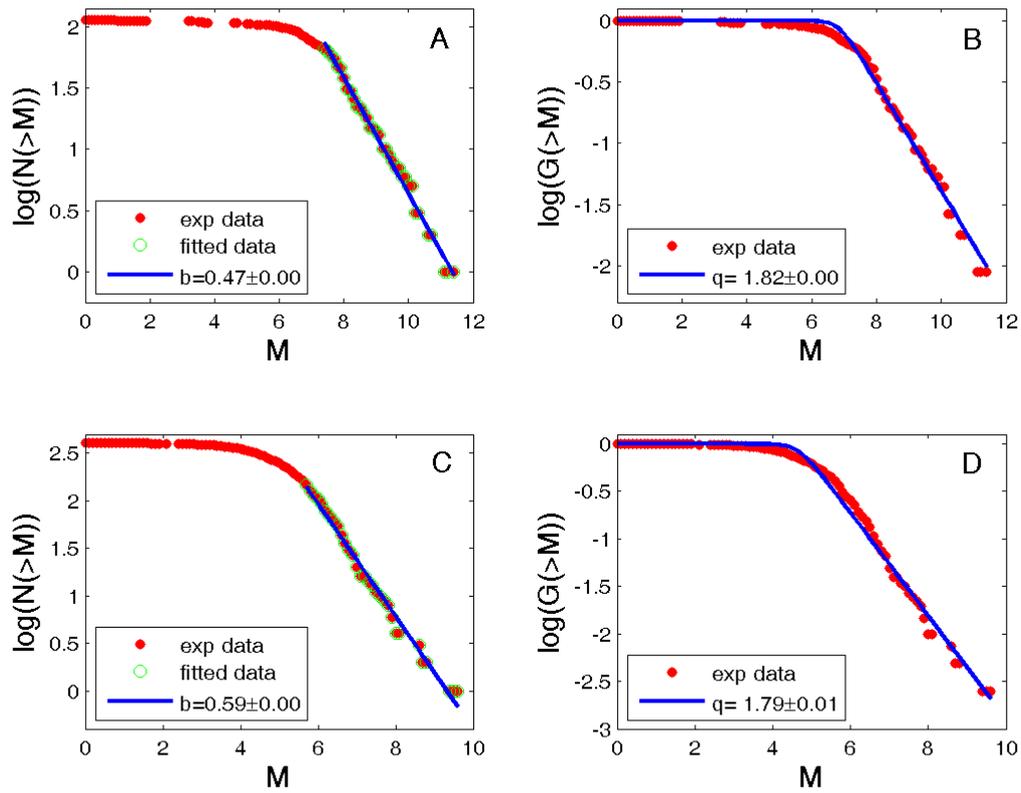

**Fig. 6.** The index S&P 500 (GSPC). A. Fitting of G-R law on the traded volume, $V-$, events; B. Fitting of the nonextensive formula of Eq. (5) on the traded volume, $V-$, events; C. Fitting of G-R law on the daily price fluctuation, $F-$, events; D. Fitting of the nonextensive formula of Eq. (5) on the daily price fluctuation, $F-$, events. The employed threshold values $A_{"noise"}$ for $V$ and $F$ were 3.10E+08 and 5.0, respectively.

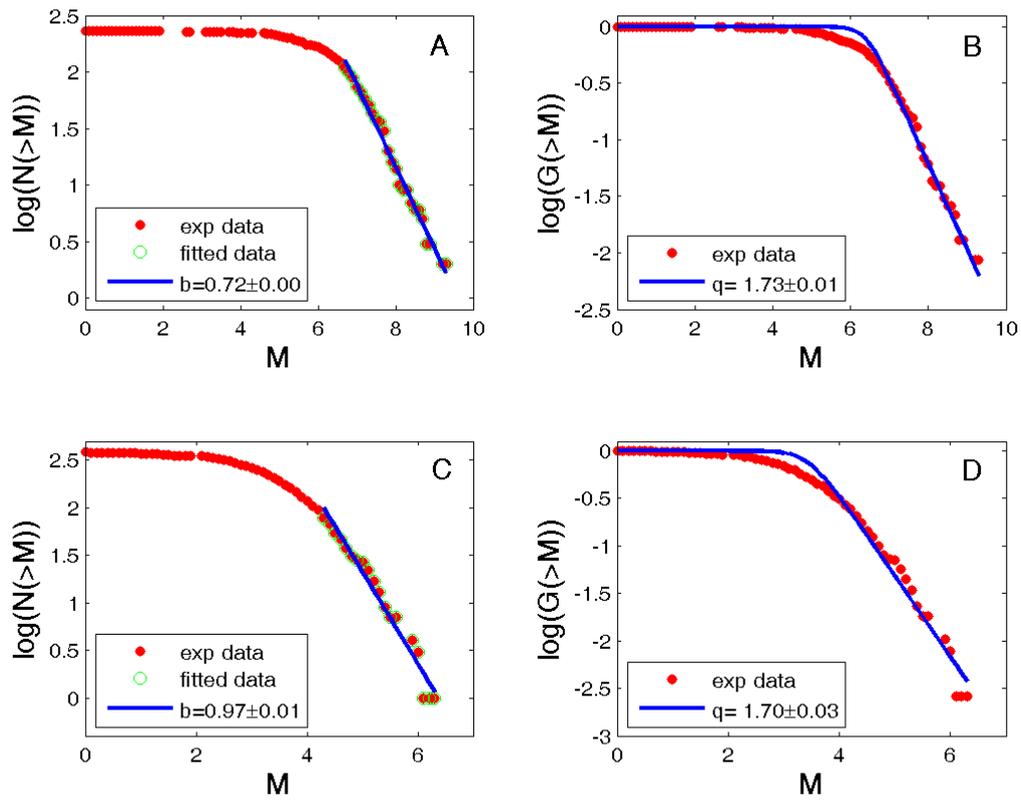

**Fig. 7.** The index SPDR Dow Jones Industrial Average (DIA). A. Fitting of G-R law on the traded volume, $V-$, events; B. Fitting of the nonextensive formula of Eq. (5) on the traded volume, $V-$, events; C. Fitting of G-R law on the daily price fluctuation, $F-$, events; D. Fitting of the nonextensive formula of Eq. (5) on the daily price fluctuation, $F-$, events. The employed threshold values $A_{"noise"}$ for $V$ and $F$ were 7.00E+06 and 1.0, respectively.

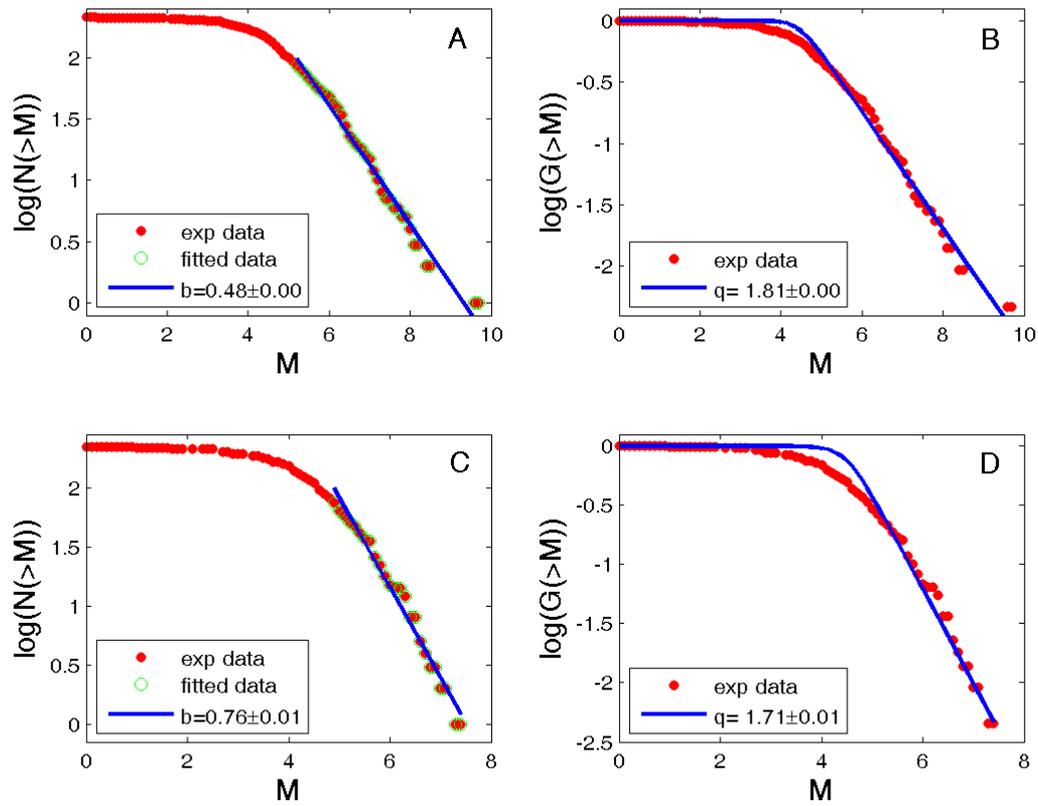

**Fig. 8.** The share United States Steel Corp. (X). A. Fitting of G-R law on the traded volume, $V-$, events; B. Fitting of the nonextensive formula of Eq. (5) on the traded volume, $V-$, events; C. Fitting of G-R law on the daily price fluctuation, $F-$, events; D. Fitting of the nonextensive formula of Eq. (5) on the daily price fluctuation, $F-$, events. The employed threshold values $A_{"noise"}$ for $V$ and $F$ were 1.00E+06 and 1.88, respectively.

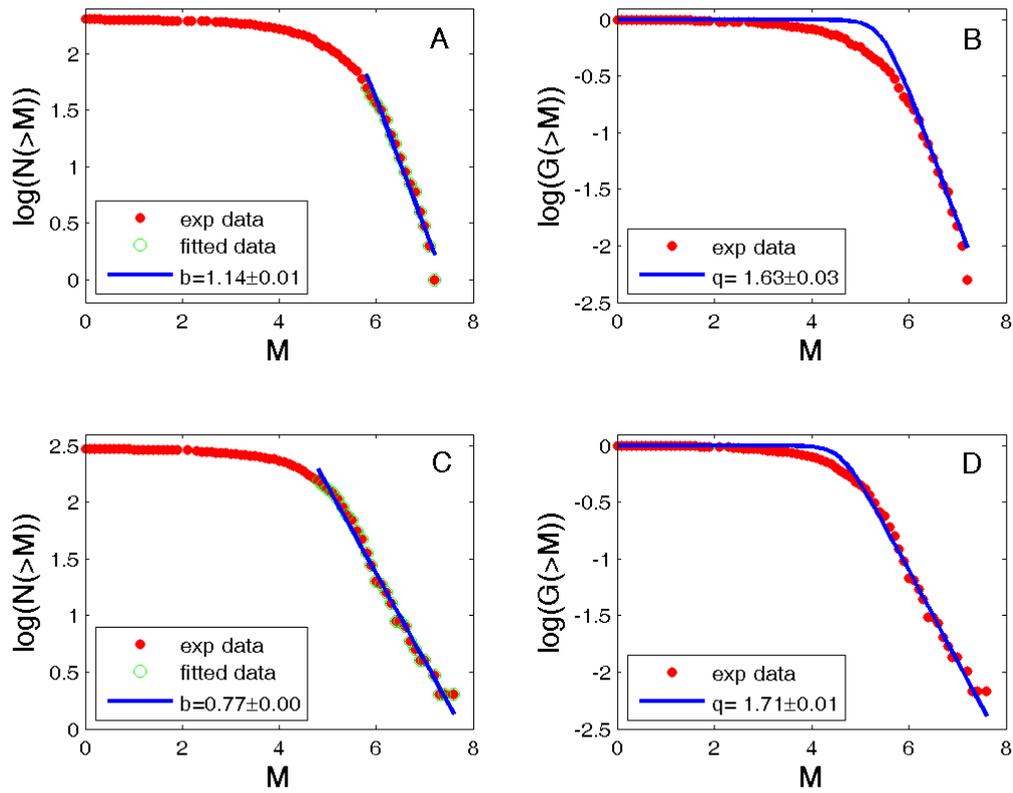

**Fig. 9.** The share Potash Corp. of Saskatchewan, Inc. (POT). A. Fitting of G-R law on the traded volume, $V-$, events; B. Fitting of the nonextensive formula of Eq. (5) on the traded volume, $V-$, events; C. Fitting of G-R law on the daily price fluctuation, $F-$, events; D. Fitting of the nonextensive formula of Eq. (5) on the daily price fluctuation, $F-$, events. The employed threshold values $A_{"noise"}$ for $V$ and $F$ were 9.00E+06 and 2.50, respectively.

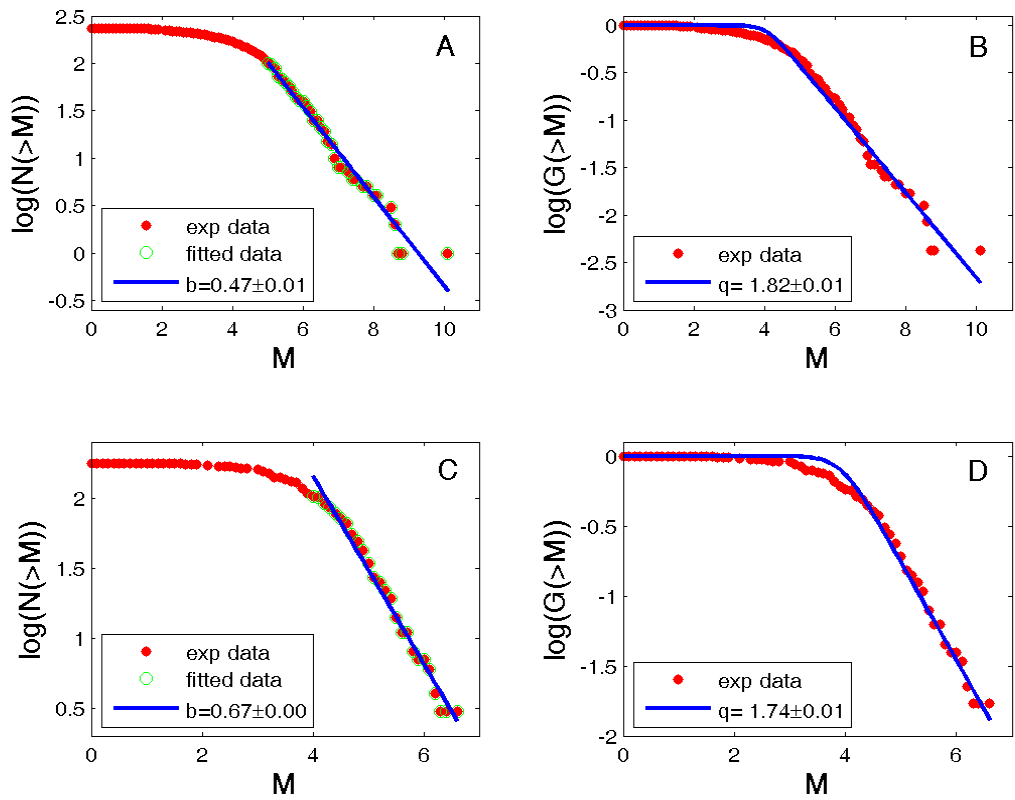

**Fig. 10.** The share Carrizo Oil & Gas Inc. (CRZO). A. Fitting of G-R law on the traded volume, $V-$, events; B. Fitting of the nonextensive formula of Eq. (5) on the traded volume, $V-$, events; C. Fitting of G-R law on the daily price fluctuation, $F-$, events; D. Fitting of the nonextensive formula of Eq. (5) on the daily price fluctuation, $F-$, events. The employed threshold values $A_{"noise"}$ for $V$ and $F$ were 2.00E+04 and 1.0, respectively.

**Table 1.** The $q-$values resulting from the fitting of Eq. (5) on the experimental data, the corresponding $b-$values from the fitting of G-R law (Eq. (4))), and the $b-$values estimated by Eq. (6), $b_{est}$, presented per share/ index along with their tolerances.

| Share / index | Event type | $q$ | $\pm \Delta q$ | $b$ | $\pm \Delta b$ | $b_{est}$ | $\pm \Delta b_{est}$ |
|---|---|---|---|---|---|---|---|
| GSPTSE | $V-$events | 1.78 | 0.02 | 0.59 | 0.01 | 0.56 | 0.07 |
|  | $F-$events | 1.74 | 0.01 | 0.79 | 0.01 | 0.70 | 0.04 |
| HSI | $V-$events | 1.75 | 0.01 | 0.65 | 0.01 | 0.67 | 0.04 |
|  | $F-$events | 1.75 | 0.01 | 0.70 | 0.00 | 0.67 | 0.04 |
| IXIC | $V-$events | 1.79 | 0.00 | 0.53 | 0.00 | 0.53 | 0.00 |
|  | $F-$events | 1.73 | 0.02 | 0.75 | 0.00 | 0.74 | 0.08 |
| GSPC | $V-$events | 1.82 | 0.00 | 0.47 | 0.00 | 0.44 | 0.00 |
|  | $F-$events | 1.79 | 0.01 | 0.59 | 0.00 | 0.53 | 0.03 |
| DIA | $V-$events | 1.73 | 0.01 | 0.72 | 0.00 | 0.74 | 0.04 |
|  | $F-$events | 1.70 | 0.03 | 0.97 | 0.01 | 0.86 | 0.12 |
| X | $V-$events | 1.81 | 0.00 | 0.48 | 0.00 | 0.47 | 0.00 |
|  | $F-$events | 1.71 | 0.01 | 0.76 | 0.01 | 0.82 | 0.04 |
| POT | $V-$events | 1.63 | 0.03 | 1.14 | 0.01 | 1.17 | 0.15 |
|  | $F-$events | 1.71 | 0.01 | 0.77 | 0.00 | 0.82 | 0.04 |
| CRZO | $V-$events | 1.82 | 0.01 | 0.47 | 0.01 | 0.44 | 0.03 |
|  | $F-$events | 1.74 | 0.01 | 0.67 | 0.00 | 0.70 | 0.04 |

*5.3. Comparison with results for EM seismogenic time-series associated to the generation of significant EQ*

It is worth noting the fact that the analysis of all the pre-crisis economic time-series yielded $q-$values lying within the range $[1.63, 1.82]$, while the resulted $b-, b_{est}-$values lie within the range $[0.44, 1.17]$, which are values compatible to the obtained from the analysis of the fracto-EM events [15-17,55-57] prior to the occurrence of a significant earthquake.

As a characteristic example, we bring attention to the results recently obtained for the seismogenic kHz EM emissions (see fig. 7 of [57]) associated with the Athens' EQ (occurred on September 7, 1999 at 11:56 (GMT) with magnitude $M_W = 5.9$). The hereby obtained results for the economic time-series are consistent with the $q-$parameters derived from the analysis of the specific kHz EM emissions, which vary between $1.78 \sim 1.82$ (see Fig. 7 of [56] and Fig. 7 of [57]). Note also that the seismicity around the epicenter of the Athens' EQ has also been analyzed with the same methods and values of $q$ for the seismicity, at close distance to the epicenter, were also found to be in the same range ($1.78 \sim 1.82$) (Fig. 8a and Fig. 8b of [57]).

In general, the $q-$values calculated for the economic time-series, further to their agreement to the ones obtained for seismogenic EM emissions, are also in agreement to the ones obtained both from the analysis of the foreshock seismicity at the geophysical scale, e.g., [47,51,57,76-80], and the analysis of foreshock acoustic emissions (AE, laboratory seismicity) at the laboratory fracture experiments scale [81-88]. Bear in mind that the herein employed definition of the magnitude of the "economic event" was given in direct analogy to the definition of the magnitude of the "fracto-EM event" or "EM-EQ". Importantly, the obtained $q-$values strongly verify the sub-extensivity of the underlying economic mechanisms, while it is in full agreement with the upper limit $q < 2$ obtained from several studies involving the study of geophysical mechanisms related to EQ generation within the Tsallis non-extensive framework [76, and references therein]. Note also that from most geological analysis performed so far, values of $q \sim 1.6 - 1.8$ seem to be universal, in the sense that different datasets from different regions of the globe indicate a value for the nonextensive parameter lying in this interval, e.g., [47,51,57,76-78].

**6. Discussion and conclusions**

As it was mentioned in introduction, many complex systems exhibit complex dynamics in which subunits of the system interact at widely varying scales of time and space. These complex interactions often generate very noisy output signals which still exhibit scale-invariant structure. An earthquake is resulted by the interaction of the opening cracks which behave as efficient electromagnetic emitters. The earthquake is a typical example of phenomena which has several temporal and spatial fractal features. Economic systems are in fact also comprised of a large number of interacting units and have the potential of displaying power law behavior. At least some economic phenomena are described by power laws [89-92, and references therein]. Pareto first investigated the statistical character of the wealth of individuals by modeling them using the scale-invariance distribution $f(x) \propto \sim x^{-a}$, where $f(x)$ denotes the number of people having income $x$ or greater than $x$, and $a$ is an exponent that Pareto estimated to be 1.5 [93]. A *twin concept of scaling was developed, i.e., the concept of universality.* Pareto noticed that his result was *universal* in the sense that it applied to populations as different as those of England, Ireland, and Germany, as well as of the Italian cities, and even of Peru. A physicist would say that the universal class of the previous power law includes all the aforementioned countries as well as Italian cities, since by definition two systems belong to the same universality class if they are characterized by the same exponents. In the century following Pareto's discovery, *the twin concepts of scaling and universality* have proved to be important in a number of systems [94,95,91]. Remarkably, the statistical properties of economic observables appear to be similar for quite different markets [96], consistent with the possibility that there may exist "universal" results. On the other hand, an earthquake nucleating anywhere at any time, will grow to a magnitude $\geq M$ according to the Gutenberg-Richter distribution (Eq. (4)). Results confirm the universality of $b-$values [97].

The new field of study of complex systems considers that the dynamics of complex systems are founded on universal principles that may be used to describe disparate problems ranging from particle physics to economies of societies [2]. This is a basic reason for our interest in complexity [98-101].

Characteristically, earthquakes and solar flares are phenomena involving huge and rapid releases of energy characterized by complex temporal occurrence. de Arcangelis et al. [4], by analyzing available experimental catalogs, have shown that the stochastic processes underlying these apparently different phenomena have universal properties. For example, both problems exhibit the same distributions of sizes. The authors conclude that observed universality suggests a common approach to the interpretation of both phenomena in terms of the same driving physical mechanism. Importantly, authors have recently studied financial fluctuations using concepts developed in the field of seismology identifying parallels between energy cascades [50-102]. The cascading dynamics immediately before and immediately after market shocks were investigated. The authors conclude that the cascade of high volatility "aftershocks" triggered by the "main shock" is quantitatively similar to earthquakes and solar flares, which have been described by three empirical laws: the Omori law, the productivity law, and the Bath law. Note that these studies refer to dynamical analogies between populations of seismic and economic shocks.

In this work we investigated, for the first time, whether the cascade of financial fluctuations associated with the generation of a *single* financial shock and pre-seismic fluctuations (in terms of the fracture induced electromagnetic emissions) connected with the generation of a *single* seismic shock are quantitatively similar. More precisely, we investigated whether the aforementioned different population of fluctuations exhibit: (i) scaling properties, and (ii) the two systems under study belong to the same universality class in the sense that they characterized by the same exponents. We proved that the result of this investigation was positive building on concepts of nonextensive statistical mechanics as it has been introduced by Tsallis and not just by the empirical Gutenberg-Richter, Omori, Bath seismological laws. More precisely, after an appropriate common definition of the magnitude of "financial" and "seismogenic electromagnetic events" (quantities exceeding certain "noise" level), we showed that the populations of: (i) fracto-EM events rooted in the activation of a single fault, (ii) the trade volume events of different shares / economic indices, leading to a price collapse, and (iii) the daily price fluctuation events of different shares / financial indices, prior to a price collapse, follow the nonextensive model for earthquake dynamics expressed by Eq. (5). It is noted that the specific nonextensive model describes in wider scale the investigated dynamics, as proven by the analysis results, since it offers a very good fitting of all magnitude values and not only of those exceeding some magnitude threshold, as it happens for the Gutenberg - Richter law (Eq. (4)). Importantly, the two different complex systems ending-up to extreme phenomena are characterized by similar non-extensive $q-$parameter values.

On the other hand, earthquake dynamics have been found to follow the empirical frequency-magnitude scaling relation, known as Gutenberg - Richter law (Eq. (4)). As it was mentioned (see Sec. 4), the $q-$parameter included in the nonextensive formula of Eq. (5), *above some magnitude threshold*, is associated with the $b-$value by Eq. (6). Either through the calculated $q-$values or by direct calculations, we also showed that all the above mentioned populations also follow the Gutenberg – Richter with similar $b-$exponents. Therefore, the obtained results prove that financial fluctuations associated with the generation of a financial main shock and pre-seismic electromagnetic fluctuations connected with the generation of a single seismic shock show scaling properties, and universality in the sense that they are characterized by the same exponents.

A question effortlessly arises whether the magnitude distribution of "events" of other different complex systems ending-up to extreme phenomena are also described by the nonextensive formula of Eq. (5) with similar $q-$parameter. Indeed, this happens in the cases of epileptic seizures, solar flares, magnetic storms, and earthquakes of a wide regional seismic region [21,26-28]. This experimental evidence enhances the suspicion that all the above mentioned various different extreme phenomena share a common dynamics in terms of nonextensive statistical mechanics. There may exist "universal" results. What may turn out to be universal is the resulting hierarchical organization, involving many scales, common to complex systems [103].

It has to be noted at this point that reservations have been expressed on the unconsidered use of results from one discipline to another, especially if they are not followed by clear theoretical basis, e.g., [104,105]. However, recent attempts to make models that reproduce the empirical scaling relationships suggest that significant progress on understanding firm growth may be under way [90]. Stanley and Plerou examine the degree to which the twin concepts of scaling and universality apply to economics systems as compared with other physical systems comprising a large number of interconnected and interacting components. Concerning the deeper comparison of seismic shocks and financial crashes, the justification for the existence of a common origin of the observed scaling similarities is needed. Scientists have reported evidence in this direction. For example, Sornette [106] has described a unified approach for modeling catastrophic events or ruptures (rupture of composite materials, great earthquakes, financial crashes, turbulence, and human birth). Bhattacharyya, Chakrabarti and colleagues [107-110] have reported arguments for the existence of a common mode of the origin of power laws in models of market and earthquakes based on the fact that the earthquake occurs due to dynamic stick-slip phenomena at the faults and the observation of the similarity in the roughness of fractured solid surfaces.

The understanding of the origin of scale invariance has been one of the fundamental tasks of modern statistical mechanics. The question of how to interpret the common power laws and associated universality remains open. The herein presented results imply similarities in scaling properties and universality between the generation of a single seismic event, and a financial crisis, but furthermore to the generation of an epileptic seizure, a magnetic storm and a solar flare. These similarities were found within a model originating from solid theoretical basis like the nonextensive statistical mechanics, which is considered appropriate for the description of systems exhibiting long-range correlations, memory, or fractal properties. Therefore, these results support the suggestion of a common approach to the interpretation of different complex extreme events in terms of the same driving mechanism.